\documentclass[twocolumn,showpacs,aps,prl,superscriptaddress,floatfix]{revtex4}
\usepackage{epsfig}

\newcommand{\BABARPubYear}{06}
\newcommand{\BABARPubNumber}{070}
\newcommand{\SLACPubNumber}{12258}
\newcommand{\LANLNumber}{0612017}

\def\effrp{11.0}                        
\def\effrz{14.1}                        
\def\effom{7.9}                         

\def\BFrp{1.10^{+0.37}_{-0.33}\pm 0.09} 
\def\BFrz{0.79^{+0.22}_{-0.20}\pm 0.06} 
\def\BFom{0.40^{+0.24}_{-0.20}\pm 0.05} 
\def\BFomUL{0.78}                       
\def\IST{-0.35 \pm 0.27} 
\def\BFav{1.25^{+0.25}_{-0.24}\pm 0.09} 
\def\BFavrhorho{1.36^{+0.29}_{-0.27}\pm0.10} 
\def\significance{6.4}                  
\def\BrhoBKst{0.030\pm0.006}            
\def\VtdVtsval{0.200 ^{+0.021} _{-0.020} \pm 0.015 } 

\def\numBB{347 million}

\def\GeV{\;\mbox{GeV}}

\def\GeVcc{\;\mbox{GeV}/c^2}

\def\MeVcc{\;\mbox{MeV}/c^2}

\def\de   {\Delta E}

\def\mes  {M_{\mbox{\scriptsize ES}} }
\def\bkg  {B \to K^{*}\gamma}

\def\brpg {B^+ \to \rho^+\gamma}
\def\brzg {B^0 \to \rho^0\gamma}
\def\bomg {B^0 \to \omega\gamma}

\def\bsg    {b\to s\gamma}

\def\rhoz {\rho^0}
\def\rhop {\rho^+}

\def\bdg    {\ensuremath {b \to d \gamma}}
\def\bkg    {\ensuremath {\B \to \Kstar \gamma}}

\def\Kz    {\ensuremath{K^{0}}\xspace}

\def\de        {\ensuremath {\Delta E}}

\def\avbr      {\ensuremath{{\BR}[B\rightarrow(\rho/\omega)\gamma]}}
\def\VtdVts    {\ensuremath{|V_{td}/V_{ts}|}}

\RequirePackage{xspace}

\usepackage{relsize}
\def\babar{\mbox{\slshape B\kern-0.1em{\smaller A}\kern-0.1em
    B\kern-0.1em{\smaller A\kern-0.2em R}}}

\def\en         {\ensuremath{e^-}\xspace}   
\def\ep         {\ensuremath{e^+}\xspace}
 
\def\epem       {\ensuremath{e^+e^-}\xspace}

\def\g     {\ensuremath{\gamma}\xspace}

\def\s     {\ensuremath{s}\xspace}

\def\b     {\ensuremath{b}\xspace}

\def\piz   {\ensuremath{\pi^0}\xspace}

\def\pip   {\ensuremath{\pi^+}\xspace}
\def\pim   {\ensuremath{\pi^-}\xspace}

\def\pipm  {\ensuremath{\pi^\pm}\xspace}

\def\Kbar  {\kern 0.2em\overline{\kern -0.2em K}{}\xspace}

\def\Kz    {\ensuremath{K^0}\xspace}
\def\Kzb   {\ensuremath{\Kbar^0}\xspace}
\def\KzKzb {\ensuremath{\Kz \kern -0.16em \Kzb}\xspace}
\def\Kp    {\ensuremath{K^+}\xspace}
\def\Km    {\ensuremath{K^-}\xspace}

\def\KpKm  {\ensuremath{\Kp \kern -0.16em \Km}\xspace}

\def\Kstar   {\ensuremath{K^*}\xspace}

\def\Dbar    {\kern 0.2em\overline{\kern -0.2em D}{}\xspace}

\def\Dz      {\ensuremath{D^0}\xspace}
\def\Dzb     {\ensuremath{\Dbar^0}\xspace}
\def\DzDzb   {\ensuremath{\Dz {\kern -0.16em \Dzb}}\xspace}
\def\Dp      {\ensuremath{D^+}\xspace}
\def\Dm      {\ensuremath{D^-}\xspace}

\def\DpDm    {\ensuremath{\Dp {\kern -0.16em \Dm}}\xspace}

\def\B       {\ensuremath{B}\xspace}
\def\Bbar    {\kern 0.18em\overline{\kern -0.18em B}{}\xspace}

\def\BB      {\ensuremath{B\Bbar}\xspace} 
\def\Bz      {\ensuremath{B^0}\xspace}
\def\Bzb     {\ensuremath{\Bbar^0}\xspace}
\def\BzBzb   {\ensuremath{\Bz {\kern -0.16em \Bzb}}\xspace}
\def\Bu      {\ensuremath{B^+}\xspace}
\def\Bub     {\ensuremath{B^-}\xspace}
\def\Bp      {\ensuremath{\Bu}\xspace}

\def\BpBm    {\ensuremath{\Bu {\kern -0.16em \Bub}}\xspace}

\def\BorBbar    {\kern 0.18em\optbar{\kern -0.18em B}{}\xspace}
\def\DorDbar    {\kern 0.18em\optbar{\kern -0.18em D}{}\xspace}
\def\KorKbar    {\kern 0.18em\optbar{\kern -0.18em K}{}\xspace}

\mathchardef\Upsilon="7107
\def\Y#1S{\ensuremath{\Upsilon{(#1S)}}\xspace}

\mathchardef\Deltares="7101
\mathchardef\Xi="7104
\mathchardef\Lambda="7103
\mathchardef\Sigma="7106
\mathchardef\Omega="710A

\def\Deltabar{\kern 0.25em\overline{\kern -0.25em \Deltares}{}\xspace}
\def\Lbar{\kern 0.2em\overline{\kern -0.2em\Lambda\kern 0.05em}\kern-0.05em{}\xspace}
\def\Sigbar{\kern 0.2em\overline{\kern -0.2em \Sigma}{}\xspace}
\def\Xibar{\kern 0.2em\overline{\kern -0.2em \Xi}{}\xspace}
\def\Obar{\kern 0.2em\overline{\kern -0.2em \Omega}{}\xspace}
\def\Nbar{\kern 0.2em\overline{\kern -0.2em N}{}\xspace}
\def\Xb{\kern 0.2em\overline{\kern -0.2em X}{}\xspace}

\def\BR         {{\ensuremath{\cal B}\xspace}}

\def\mes        {\mbox{$m_{\rm ES}$}\xspace}

\def\DeltaE     {\mbox{$\Delta E$}\xspace}

\newcommand{\tev}{\ensuremath{\mathrm{\,Te\kern -0.1em V}}\xspace}
\newcommand{\gev}{\ensuremath{\mathrm{\,Ge\kern -0.1em V}}\xspace}
\newcommand{\mev}{\ensuremath{\mathrm{\,Me\kern -0.1em V}}\xspace}
\newcommand{\kev}{\ensuremath{\mathrm{\,ke\kern -0.1em V}}\xspace}
\newcommand{\ev}{\ensuremath{\mathrm{\,e\kern -0.1em V}}\xspace}
\newcommand{\gevc}{\ensuremath{{\mathrm{\,Ge\kern -0.1em V\!/}c}}\xspace}
\newcommand{\mevc}{\ensuremath{{\mathrm{\,Me\kern -0.1em V\!/}c}}\xspace}
\newcommand{\gevcc}{\ensuremath{{\mathrm{\,Ge\kern -0.1em V\!/}c^2}}\xspace}
\newcommand{\mevcc}{\ensuremath{{\mathrm{\,Me\kern -0.1em V\!/}c^2}}\xspace}

\def\mus  {\ensuremath{\rm \,\mus}\xspace}

\def\mus        {\ensuremath{\,\mu{\rm s}}\xspace}    

\def\calL{{\ensuremath{\cal L}}\xspace}

\def\to                 {\ensuremath{\rightarrow}\xspace}

\def\pep2{PEP-II}

\def\gsim{{~\raise.15em\hbox{$>$}\kern-.85em
          \lower.35em\hbox{$\sim$}~}\xspace}
\def\lsim{{~\raise.15em\hbox{$<$}\kern-.85em
          \lower.35em\hbox{$\sim$}~}\xspace}

\xspace

\newcommand{\epjBase}        {Eur.\ Phys.\ Jour.\xspace}
\newcommand{\jprlBase}       {Phys.\ Rev.\ Lett.\xspace}
\newcommand{\jprBase}        {Phys.\ Rev.\xspace}
\newcommand{\jplBase}        {Phys.\ Lett.\xspace}

\newcommand{\npBase}         {Nucl.\ Phys.\xspace}

\newcommand{\epjc}      [1]  {\epjBase\ C~{\bf #1}}

\newcommand{\npb}       [1]  {\npBase\ B~{\bf #1}}

\newcommand{\plb}       [1]  {\jplBase\ B~{\bf #1}}

\newcommand{\jprl}      [1]  {\jprlBase\ {\bf #1}}
\newcommand{\jprd}      [1]  {\jprBase\ D~{\bf #1}}

\def\jetset74   {\mbox{\tt Jetset \hspace{-0.5em}7.\hspace{-0.2em}4}\xspace}

\begin{document}

\preprint{\babar-PUB-\BABARPubYear/\BABARPubNumber}
\preprint{SLAC-PUB-\SLACPubNumber}

\begin{flushleft}
\babar-PUB-\BABARPubYear/\BABARPubNumber\\
SLAC-PUB-\SLACPubNumber\\
hep-ex/\LANLNumber\\
\end{flushleft}

\begin{flushright}
\end{flushright}
\title{\large\bf\boldmath
    Branching Fraction Measurements of
    $B^+\to \rho^+\gamma$, $B^0\to \rho^0\gamma$, and $B^0\to\omega\gamma$
}

%
\author{B.~Aubert}
\author{M.~Bona}
\author{D.~Boutigny}
\author{Y.~Karyotakis}
\author{J.~P.~Lees}
\author{V.~Poireau}
\author{X.~Prudent}
\author{V.~Tisserand}
\author{A.~Zghiche}
\affiliation{Laboratoire de Physique des Particules, IN2P3/CNRS et Universit\'e de Savoie, F-74941 Annecy-Le-Vieux, France }
\author{E.~Grauges}
\affiliation{Universitat de Barcelona, Facultat de Fisica, Departament ECM, E-08028 Barcelona, Spain }
% ...
\collaboration{The \babar\ Collaboration}
\noaffiliation

\date{\today}

\begin{abstract}
We present a study of  the decays
$B^+\to\rho^+\gamma$, $B^0\to \rho^0\gamma$, and $B^0\to\omega\gamma$.
The analysis is based on data containing \numBB\ \BB\
events recorded with
the \babar\ detector at the PEP-II asymmetric $B$ factory.
We measure the branching fractions
$\BR(\Bp\to\rhop\gamma) = (\BFrp)\times10^{-6}$
and
$\BR(\Bz\to\rhoz\gamma) = (\BFrz)\times10^{-6}$,
and set a 90\% C.L.\ upper limit
$\BR(\Bz\to\omega\gamma) < \BFomUL\times10^{-6}$.
We also measure the isospin-averaged branching fraction
$ \avbr
= (\BFav) \times10^{-6}$,
from which we determine
$\VtdVts = \VtdVtsval$,
where the first uncertainty is experimental and the second is theoretical.
\end{abstract}


\maketitle

In the Standard Model,  the decays $\brpg$, $\brzg$,
and $\bomg$\cite{ChargeConj} arise mainly from $\bdg$
penguin diagrams containing a virtual top quark in the loop.
By relating the three individual decay rates by isospin symmetry and
using the measured ratio between the charged and neutral $B$ meson
lifetimes $\tau_{\Bp}/\tau_{\Bz}$, an isospin-averaged branching
fraction is defined:
$    \avbr \equiv \frac{1}{2} \left\{
         \BR(\brpg) + \frac{\tau_{\Bp}}{\tau_{\Bz}}\left[
             \BR(\brzg) + \BR(\bomg)
         \right]
    \right\}.
$
Recent calculations predict $\avbr$ to be in the range of
$(0.9$--$1.8)\times 10^{-6}$~\cite{ali2004,SM},
where most of the uncertainty is due to the calculation of the form factor.
These predictions could
be modified by processes beyond the Standard Model~\cite{hewett}.

While the exclusive decay rates have a large uncertainty due to non-perturbative
long-distance QCD effects, some of this uncertainty cancels in the ratio
of
\hbox{
\avbr\
 to $B$ \to $K^{*}$\g} branching fractions.
Since the dominant  diagram
involves a virtual top quark, this ratio
is related to the ratio of Cabibbo-Kobayashi-Maskawa (CKM) matrix elements
$\VtdVts$~\cite{alivtdvtstheory,ali2004} via
\begin{equation}\label{eq:vtd}
\frac{\avbr}
{{\cal B}(B \rightarrow K^{*}\gamma)}=
\left| \frac{V_{td}}{V_{ts}} \right|^{2}
\left(\frac{1-m_{\rho}^{2}/M_{B}^{2}}{1-m_{K^{*}}^{2}/M_{B}^{2}}\right)^{3}
\zeta^{2} [1+\Delta R].
\end{equation}
The coefficient $\zeta$ is the ratio of the form factors for the decays $B \rightarrow \rho\gamma $
and $B \rightarrow K^{*}\gamma$
and
 $\Delta R$ accounts for different dynamics in the decay
({\it e.g.} annihilation diagrams can contribute to $\brpg$).
Physics beyond the Standard Model
could affect these decays, creating inconsistencies
between the measurement of $\VtdVts$ obtained from this analysis and that obtained
from the ratio of $B^0$ and $B^0_s$ mixing frequencies~\cite{bsmixing}.

Previous searches by $\babar$~\cite{oldbabar} and CLEO \cite{cleo}
found no evidence for the decays
{\ensuremath{B\rightarrow\rho\gamma}}
and {\ensuremath{B\rightarrow\omega\gamma}}.
An observation of the decay $\brzg$ was recently reported by the Belle
collaboration~\cite{newbelle}.
This letter reports on a study of the decays
$\Bp\to\rhop\gamma$, $\Bz\to\rhoz\gamma$, and $\Bz\to\omega\gamma$
based on a data sample containing \numBB\ \BB\ 
events, corresponding to an integrated luminosity of 316 fb$^{-1}$, collected with
the \babar\ detector~\cite{ref:babar} at the \pep2\ asymmetric--energy $\epem$ storage ring.
These results supersede the previous \babar\ measurements~\cite{oldbabar}.

The decays
{\ensuremath{B\rightarrow\rho\gamma}} and {\ensuremath{B\rightarrow\omega\gamma}}
 are reconstructed by combining a high-energy photon with a
vector meson  reconstructed in the decay modes
$\rho^0\to\pip\pim$ ($\BR\sim100\%$),
$\rho^+\to\pip\piz$ ($\BR\sim100\%$),
and $\omega\to\pip\pim\piz$ ($\BR=[89.1\pm0.7]\%$)~\cite{pdg}.

The dominant source of background is continuum events
($\ep\en \to q\bar{q}$, with $q=u,d,s,c$) that contain a
high-energy photon from $\piz$ or $\eta$ decays.
Other backgrounds include
photons from initial-state radiation (ISR) processes,
decays of $\bkg$ ($\Kstar\rightarrow K\pi$),
decays of $\B\rightarrow(\rho/\omega)\piz$ or $\B\rightarrow(\rho/\omega)\eta$
and combinatorial background  from higher-multiplicity $\b\rightarrow\s\gamma$ decays.
For each signal decay mode, 
selection requirements have been optimized
for maximum statistical sensitivity with assumed  signal branching fractions
of $1.0\times 10^{-6}$ and $0.5\times 10^{-6}$  for the charged and neutral 
modes, respectively.

The photon from a signal $B$ decay is identified as a  well-isolated
 energy deposit in the electromagnetic calorimeter
with energy $1.5 <E^*_\gamma <3.5\GeV$ in the  center of mass (CM) frame.
The energy deposit must not be associated with any  charged track
and must meet several other
requirements designed to eliminate background from
hadronic showers and charged particles~\cite{babarksg}.
 In order to veto photons
  from $\piz$ and $\eta$ decays, we associate each high-energy photon candidate
  $\gamma$ with each of the other photons $\gamma'$ in the event. 
We reject the candidates that are consistent with originating from $\pi^0$ 
or $\eta$ decays based on a likelihood ratio constructed from 
the energy of the second photon $\gamma'$ 
 and 
the invariant mass of the pair  $m_{\gamma\gamma'}$.
We also combine the high-energy photon candidate with photon conversions
to $\epem$ pairs, and reject the photon if the invariant mass is consistent
with a $\piz$ or $\eta$.

Charged-pion candidates are selected from well-reconstructed tracks with a
minimum momentum transverse to the beam direction of $100~\mevc$.
A stringent $\pipm$ selection algorithm  \cite{oldbabar}
is applied to reduce background
from charged kaons produced in $\bsg$ decays.
The algorithm combines the
information provided by the ring-imaging Cherenkov detector
with the measurement of energy loss in the tracking system.

Photon candidates  with energy greater than $50\mev$ in the laboratory frame
are combined into pairs to form $\piz$ candidates. For $\Bz\rightarrow\omega\gamma$
$(\brpg)$ decays, the invariant mass of the pair is required to satisfy
$122 < m_{\gamma\gamma} < 150\MeVcc$ $(117 < m_{\gamma\gamma} < 148\MeVcc)$. We
also require that the cosine of the opening angle between the daughter photons in the
laboratory frame be greater than 0.413 (0.789). %

The identified pions are combined into vector meson candidates by requiring
$633 < m_{\pip\pim} < 957\MeVcc$, $636 < m_{\pip\piz} < 932\MeVcc$, and
$764 < m_{\pip\pim\piz} < 795\MeVcc$ for $\rho^0$, $\rho^+$,
and $\omega$, respectively.
The charged pion pairs must originate from a common vertex.
The separation along the beam axis between this vertex and the one
obtained by combining the other  charged particles in the event is
required to be less than 4~mm and to be measured with
a precision better than  0.4~mm.

The photon and $\rho/\omega$ candidates are combined to form the $B$ meson
candidates. We define $\de \equiv E^*_{B}-E_{\rm beam}^*$, where $E^*_B$ is
the CM energy of the $B$ meson candidate and
$E_{\rm beam}^*$ is the CM beam energy.
We also define the beam-energy-substituted mass $\mes \equiv
\sqrt{ E^{*2}_{\rm beam}-{\mathrm{{p}}}_{B}^{\;*2}}$, where ${\mathrm{{p}}}_B^{\;*}$ is
the CM momentum of the $B$ candidate.
Signal events are expected to have a $\de$ distribution centered near 
zero with a resolution of about $50\mev$,  and an $\mes$ distribution 
centered at the mass of the $B$ meson, with a resolution of $3~\mevcc$.
We consider
candidates in the ranges $-0.3 < \de <0.3 \GeV$ and $\mes >  5.22\GeVcc$,
which include sidebands that allow the combinatorial background yields
to be extracted from a fit to the data.

In signal events the vector meson is transversely polarized,  
while in background events from 
$\B\rightarrow\rho(\piz/\eta)$ and $\B\rightarrow\omega(\piz/\eta)$  
it is longitudinally polarized. 
To reject this background, we calculate the vector meson helicity angle,
$\theta_H$, and require $|\cos\theta_{H}|< 0.75$.
The helicity angle is defined as the angle between the $B$ momentum vector
and the $\pi^-$ track calculated in the $\rho$ rest frame in the case of a $\rho$ meson,
or the angle between the $B$ momentum vector and the normal to the $\omega$ decay plane
for an $\omega$ meson.

Contributions from continuum background processes are reduced by considering
only events for which the ratio $R_2$ of second-to-zeroth Fox-Wolfram
moments~\cite{fox} calculated using the momenta of all charged and neutral particles in the event is less than 0.7.
A neural network combining the variables described below further suppresses
continuum background.
The quantity $R'_2$,
defined as  $R_2$ in the frame recoiling against the photon momentum, is used to reject
ISR events. To discriminate between the
jet-like continuum background and the more spherically-symmetric signal
events, we compute the angle between the photon and the thrust axis of the
rest of the event (ROE) in the CM frame. The ROE is defined by all
 charged tracks and neutral energy deposits in the calorimeter that are not
used to reconstruct the $B$ candidate.
We also calculate the moments $L_{i} \equiv \sum_{j} p^{*}_{j}
\cdot|\cos{\theta^{*}_{j}}|^{i}/\sum_{j} p^{*}_{j}$, where $p^{*}_j$ and
$\theta^{*}_{j}$ are the momentum and angle with respect to an axis,
respectively, for each particle $j$ in the ROE. We use $L_{1}$, $L_{2}$,
and $L_{3}$ with respect to the thrust axis of the ROE, as well as
with respect to the photon direction. In addition, we calculate the $B$ meson
production angle $\theta_B^*$ with respect to the beam axis in the CM frame.
Differences in lepton and kaon production between background and \B
decays are exploited by using flavor-tagging variables~\cite{babartag}.
The significance of the separation
along the beam axis of the $B$ meson candidate and ROE vertices is included
as well.
The purity of the selected sample is enhanced by a cut on the output of the neural network that
retains 63\%, 74\%, and 71\% of the signal events in the modes
\mbox{$\brpg$}, \mbox{$\brzg$}, and \mbox{$\bomg$}, respectively.

The expected average candidate multiplicity in the selected signal events
is 1.01 for $\brzg$ and 1.07 for both $\brpg$ and $\bomg$. In events with multiple candidates, 
the one with the reconstructed vector meson mass closest to the nominal mass is retained.
This criteria was chosen because the mass of the vector meson was 
found to be uncorrelated with the variables used in the fit.
Applying all the selection criteria described above, we find efficiencies~\cite{effcomment}
of \effrp\% for $\brpg$, \effrz\% for $\brzg$, and \effom\% for $\bomg$.

The signal content of the data is determined by a multidimensional
unbinned maximum likelihood fit, which is constructed individually for each of the
three signal decay modes. All fits use \DeltaE , $\mes$,  $\cos\theta_H$, and
the neural network output $N$. In order
to facilitate the parametrization of the probability density function
(PDF) used in the fit, the transformation
${\cal NN} = \tanh^{-1} \left( c_1\cdot N - c_2 \right)$,
in which the $c_i$ are are mode-dependent constants, is made.
For decays $\bomg$ ($\omega\to\pi^+\pi^-\pi^0$), the cosine of the angle between the \pip and \piz momenta
in the $\pi^+\pi^-$ rest frame (Dalitz angle)  is added as a fifth observable.

In the fit we consider several hypotheses for the origin of the events:
signal,  continuum background,   $\bkg$ decays, and other $B$ backgrounds.
The likelihood
 function for a  signal mode $k$ ($=\rhop\gamma$, $\rhoz\gamma$, $\omega\gamma$) is defined as
  \begin{equation}
  {\cal L}_{k}=\exp{\left(-\sum_{i=1}^{N_{\mathrm{hyp}}} n_{i}\right)}
  \left
  [\prod_{j = 1}^{N_k}\left(\sum_{i=1}^{N_{\mathrm{hyp}}} n_i{\cal P}_{i}(\vec{x}_j;\vec{\alpha}_i)\right)\right],
  \label{LH}
  \end{equation}
 where  $N_{\mathrm{hyp}}$ is the number of event hypotheses,
 $n_i$ is the yield of each hypothesis, and $N_k$ is the number of candidate
 events observed in data.
 Since the  correlations among the
 observables are found to be small in simulated event samples,
 we  define the PDF
 $\mathcal{P}_{i}(\vec{x_j}; \vec{\alpha_{i}})$ for the $i$-th event hypothesis
 as the product of individual PDFs for each fit observable
 ${x}_{j}$  given the set of parameters $\vec{\alpha}_{i}$.

Each PDF is determined from a one-dimensional fit to a dedicated
sample of simulated events.
The $\de$ PDF is corrected for the observed difference between data
and simulation  by using samples of  $\bkg$ decays.
All continuum background
parameters
float freely in the fits, while the shapes of the signal and $B$ background
distributions are fixed according to the Monte Carlo simulation.
The signal $\mes$ spectra are described by Crystal Ball functions \cite{CryBall},
the angular distributions are modeled by
polynomials,
and the distributions of $\de$ and ${\cal NN}$ are parametrized as
$f(x) = \exp \left( \frac{-(x-\mu)^2}{2 \sigma^2_{L,R} + \alpha_{L,R} (x-\mu)^2} \right)$,
where $\mu$ is the peak position of the distribution, $\sigma_{L,R}$ are the
widths on the left and right
of the peak, and $\alpha_{L,R}$ are a measure of the
tails on the left and right of the peak,
respectively.
Various functional forms are used to describe the continuum and $B$ background components.

\begin{figure}
\includegraphics[width=0.5\linewidth,clip=true]{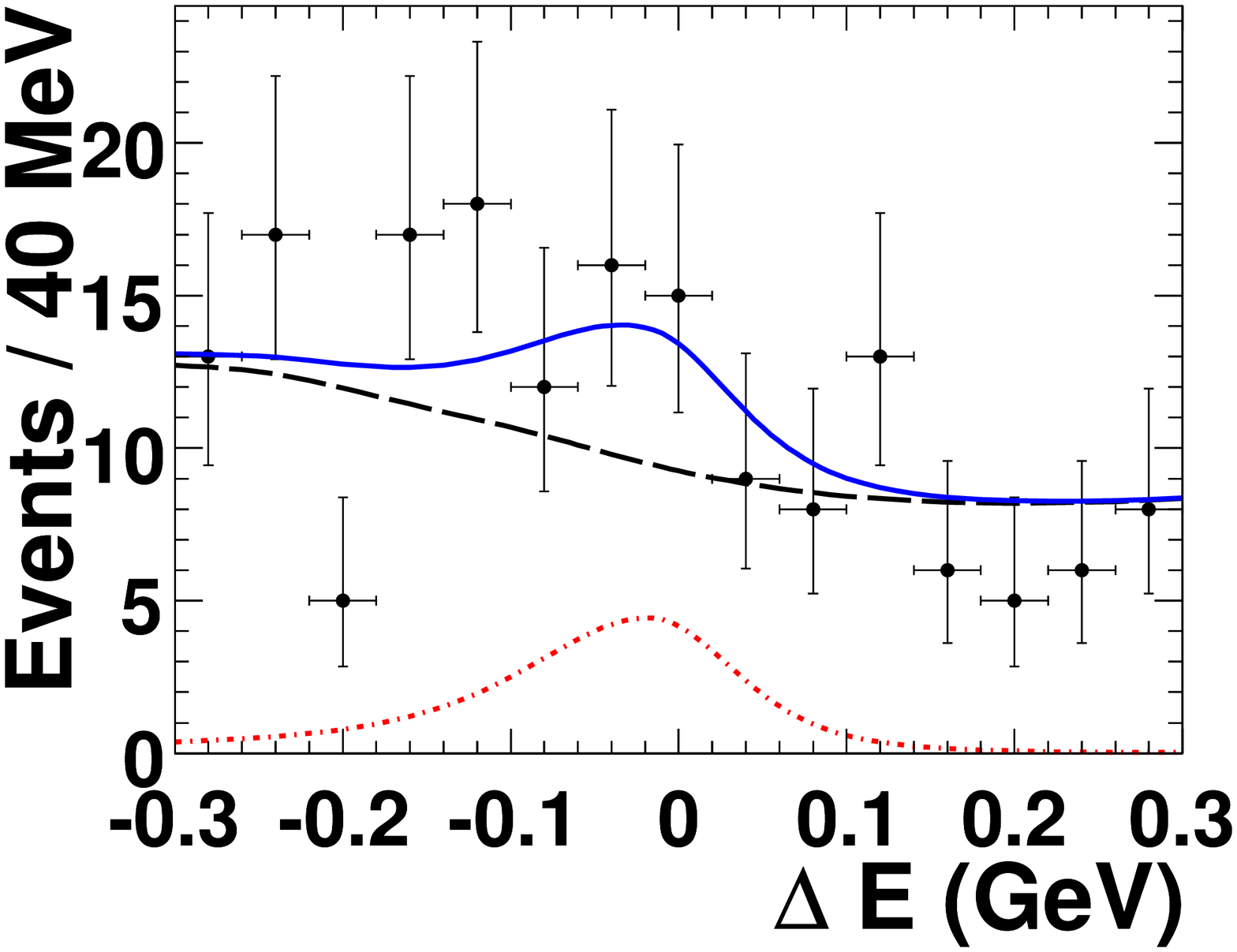}%
\includegraphics[width=0.5\linewidth,clip=true]{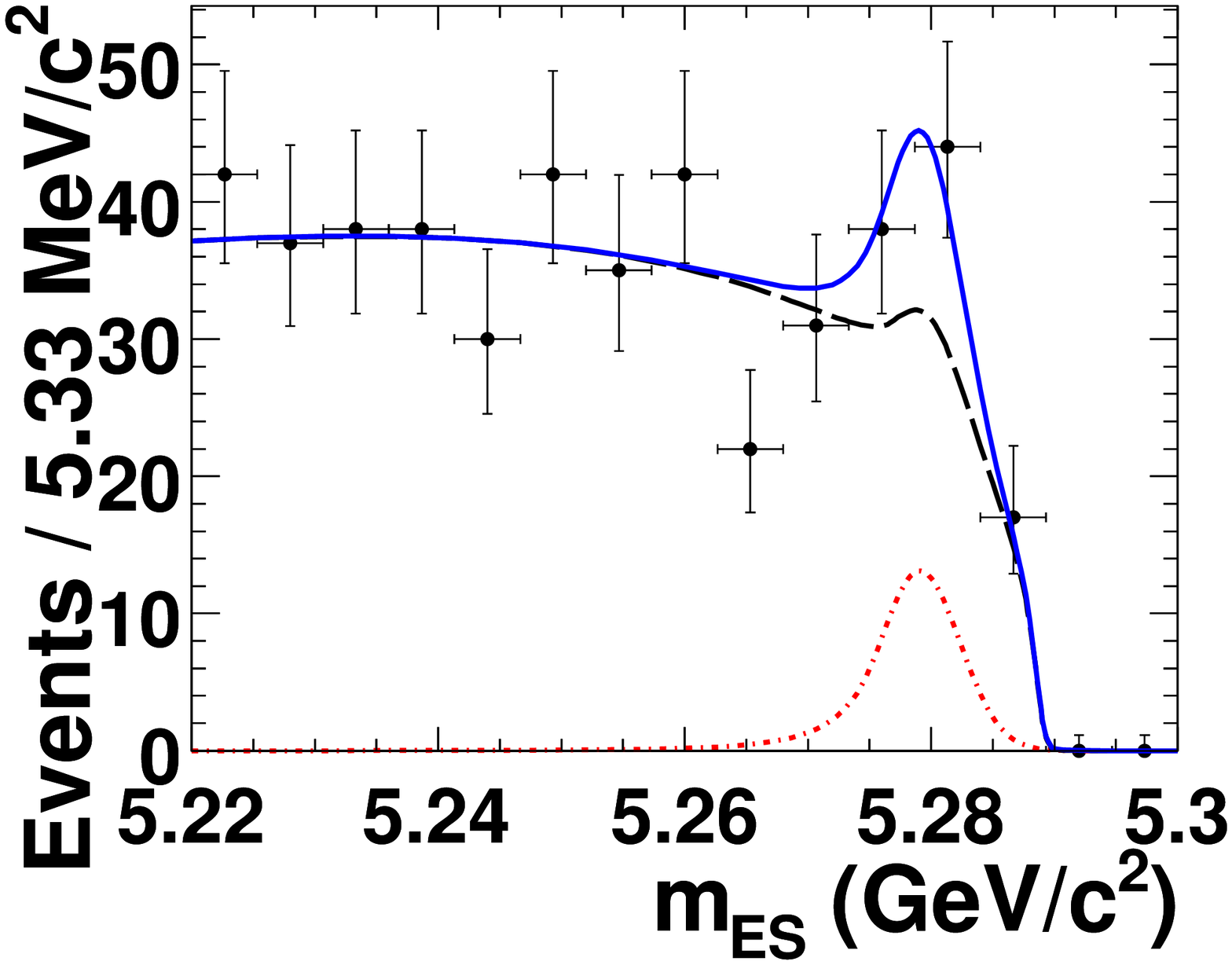}\\
\includegraphics[width=0.5\linewidth,clip=true]{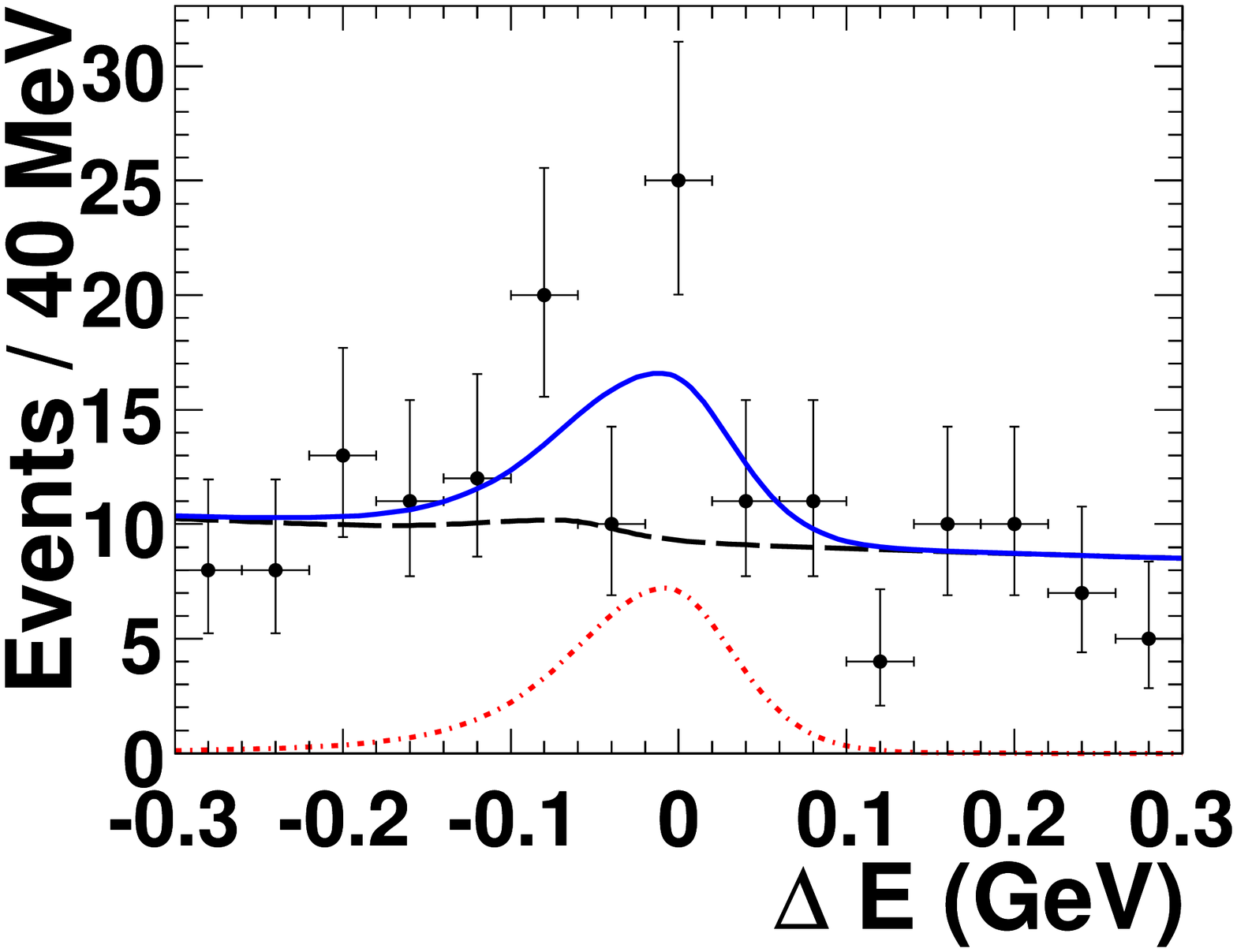}%
\includegraphics[width=0.5\linewidth,clip=true]{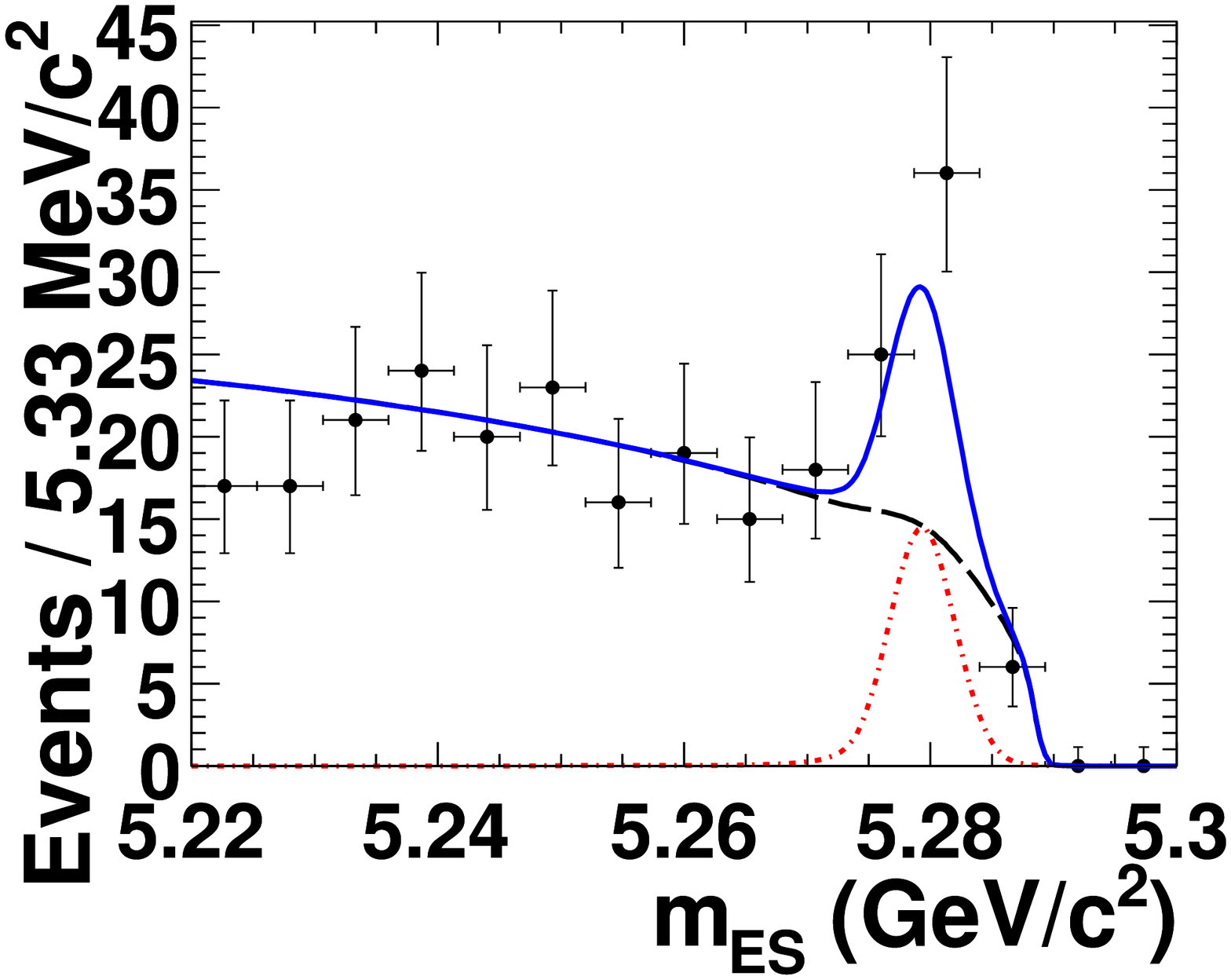}\\
\includegraphics[width=0.5\linewidth,clip=true]{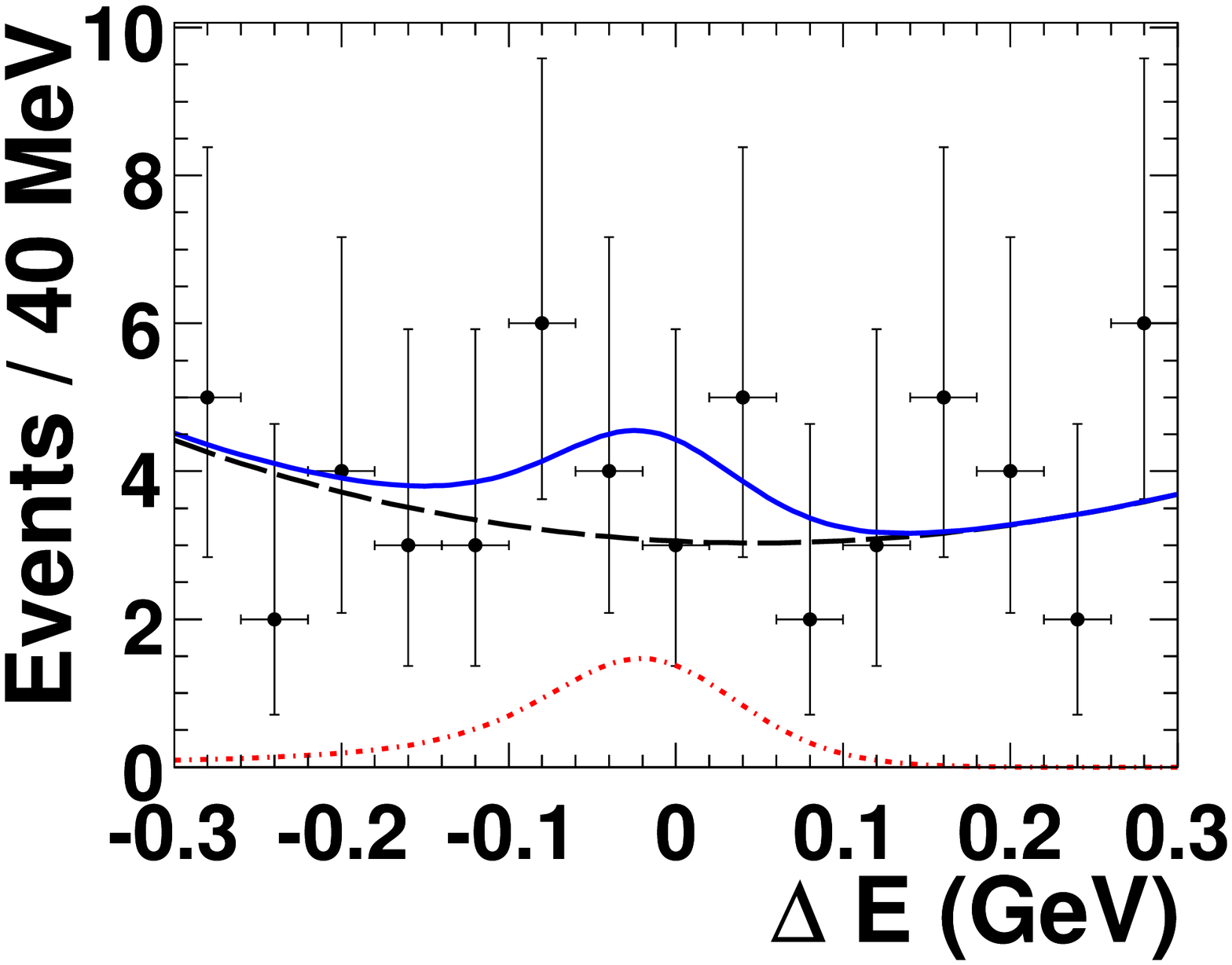}%
\includegraphics[width=0.5\linewidth,clip=true]{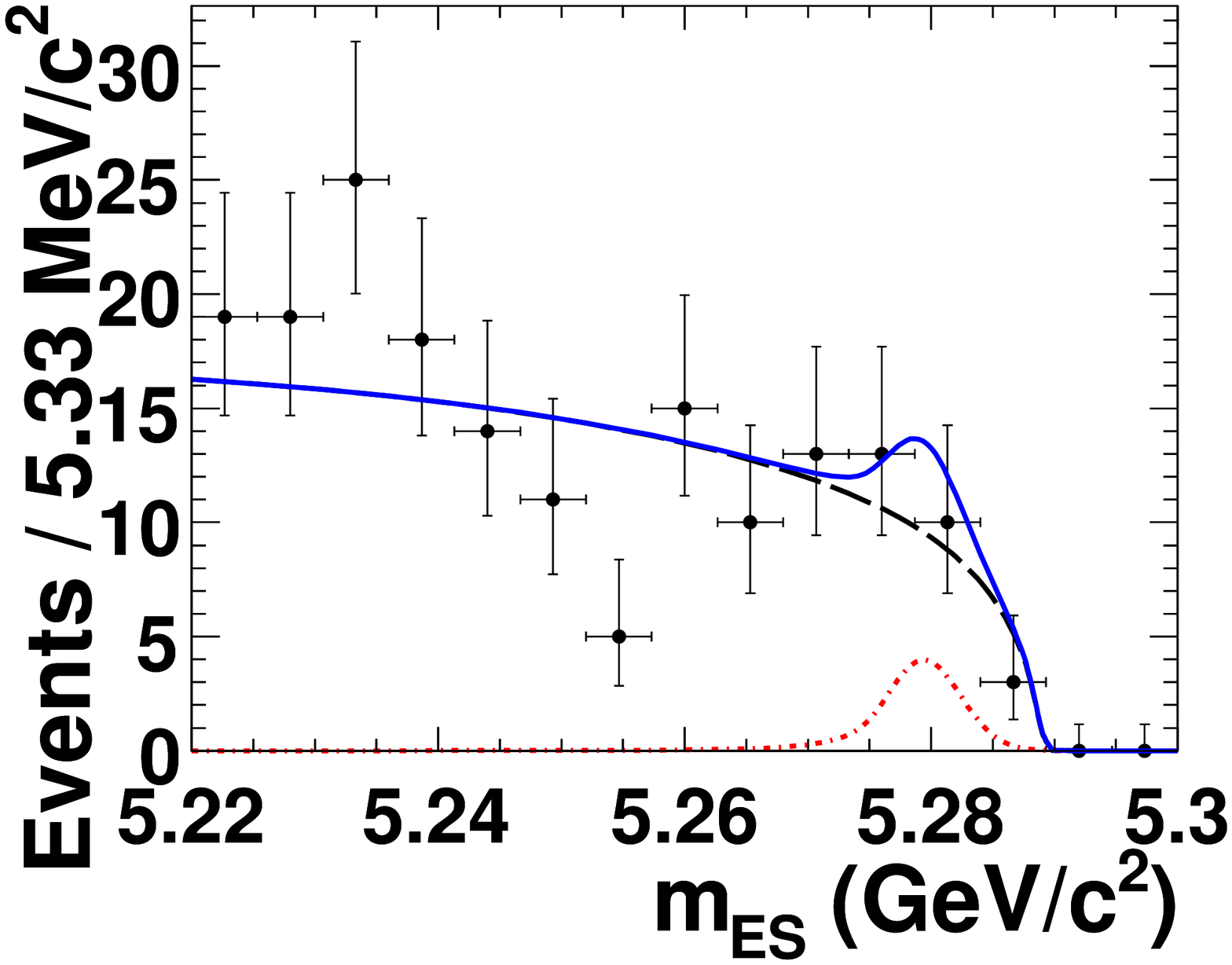}\\
\caption{
$\de$ and $\mes$ projections of the fits for the decay modes $\brpg$ (top),
$\brzg$ (middle), and $\bomg$ (bottom).
In each plot, the signal fraction is enhanced by selections  on the other fit variables.
The points are data, the solid line is the total of all contributions, and
the long-dashed (dashed-dotted) line is background-only (signal-only).
}\label{fig:rhoChfit}
\end{figure}

We measure the signal yield $n_{sig}$ by maximizing the likelihood function in Eq.~(\ref{LH}).
In the fit, the continuum background yield is allowed to float, as is the overall yield
of the $B$ background, with the exception of the $B^+ \to K^{*+} \gamma$
($K^{*+} \to K^{+} \pi^0$) yield in the $\brpg$  mode, which is fixed.
The relative yields among the different $B$ backgrounds are fixed to the values obtained
 using known branching fractions~\cite{pdg} and
selection efficiencies determined from simulated events.
Figure \ref{fig:rhoChfit} shows the data points and the projections of the fit
results for \DeltaE and  $\mes$  separately for each decay mode.
The  signal yields are reported in
Table~\ref{tab:results}. The significance is computed as
$\sqrt{2\Delta\log\mathcal{L}}$, where $\Delta\log\mathcal{L}$ is the
log-likelihood difference between the best fit and the null-signal hypothesis.
%

Table~\ref{tab:syst} gives an overview of the contributions
to the systematic uncertainties. These are associated with the signal
reconstruction efficiency and the modeling of signal and $\BB$ background
in the Monte Carlo simulation.
The latter contributes  to the uncertainties on the signal yields.
\begin{table}
\centering
\caption{\label{tab:results} The signal yield $(n_{\mathrm{sig}})$,
significance ($\Sigma$) in standard deviations including systematic errors, efficiency $(\epsilon)$,
and branching fraction $(\mathcal{B})$ for each mode.
The errors on $n_{\mathrm{sig}}$ are statistical only, while for the branching fraction
the first error is statistical and the second systematic.}
\renewcommand{\arraystretch}{1.3}
\begin{tabular}{l@{\hspace{0.4cm}}c@{\hspace{0.4cm}}c@{\hspace{0.4cm}}r@{\hspace{0.4cm}}c@{\hspace{0.4cm}}c}
\hline
\hline
Mode & $n_{\mathrm{sig}}$ & $\Sigma$
     & $\epsilon (\%)$    & $\mathcal{\B} (10^{-6})$ \\
\hline
$\brpg$  & $42.0^{+14.0}_{-12.7}$     & $3.8\sigma$ &  $\effrp$ & $\BFrp$ \\
$\brzg$  & $38.7^{+10.6}_{-9.8}$      & $4.9\sigma$ &  $\effrz$ & $\BFrz$ \\
$\bomg$  & $11.0^{+6.7}_{-5.6}$       & $2.2\sigma$ &  $\effom$ & $\BFom$ \\
\hline
$ B\rightarrow(\rho/\omega)\gamma $ & & $\significance\sigma$ &  & $\BFav $ \\
\hline
$ B\rightarrow \rho\gamma $ & & $6.0\sigma$ &  & $\BFavrhorho $ \\
\hline
\hline
\end{tabular}
\end{table}
\begin{table}
\renewcommand{\arraystretch}{1.3}
\centering \caption{\label{tab:syst}
Fractional systematic errors (in \%) of the measured branching fractions.}
\begin{tabular*}{\linewidth}{lp{0.9cm}p{0.9cm}p{0.9cm}p{0.65cm}p{0.9cm}}
\hline \hline
\multicolumn{1}{p{3.5cm}}{Source of error}
&  $\rho^{+}\gamma$  &  $\rho^{0}\gamma$
&  $\omega\gamma$  & $\rho\gamma$ &  $(\rho/\omega)\gamma$    \\
\hline
Tracking efficiency             &      1.0  &  2.0  &  2.0  &  1.4  &  \multicolumn{1}{c}{1.5}  \\ 
Particle identification         &      2.0  &  4.0  &  2.0  &  2.9  &  \multicolumn{1}{c}{2.7}  \\
Photon selection                &      1.9  &  2.6  &  1.7  &  2.2  &  \multicolumn{1}{c}{2.1}  \\ 
$\pi^0$ reconstruction          &      3.0  &  -    &  3.0  &  1.9  &  \multicolumn{1}{c}{2.5}  \\
$\pi^0$ and $\eta$ veto         &      2.8  &  2.8  &  2.8  &  2.8  &  \multicolumn{1}{c}{2.8}  \\
$\mathcal{NN}$ efficiency       &      1.0  &  1.0  &  1.0  &  1.0  &  \multicolumn{1}{c}{1.0}  \\
$\mathcal{NN}$ shape            &      0.4  &  0.3  &  2.3  &  0.4  &  \multicolumn{1}{c}{0.7}  \\
Signal PDF shapes               &      4.8  &  3.3  &  2.4  &  3.1  &  \multicolumn{1}{c}{2.6}  \\
$B$ background PDFs             &      3.9  &  2.9  &  9.7  &  3.2  &  \multicolumn{1}{c}{3.1}  \\
$B\overline{B}$ sample size     &      1.1  &  1.1  &  1.1  &  1.1  &  \multicolumn{1}{c}{1.1}  \\
\BR($\omega\to\pi^+\pi^-\pi^0$) &      -    &  -    &  0.8  &  -    &  \multicolumn{1}{c}{0.1}  \\
\hline
Sum in quadrature               &      8.1  &  7.4  &  11.6 &  7.0  &  \multicolumn{1}{c}{6.9}  \\
\hline \hline
\end{tabular*}
\end{table}
The systematic error affecting the signal efficiency
includes uncertainties on tracking, particle identification, \g and \piz reconstruction,
\piz/$\eta$ veto and the neural network selection. The uncertainties  
on the \piz/$\eta$ veto and neural network selection 
 are determined from a control sample of $B\to D\pi$ decays, with $D\to K\pi$ or $D\to K\pi\pi$.  
To estimate the uncertainty related to
the modeling of the signal and $B$ background in the Monte Carlo,
we vary the parameters of the PDFs that are fixed in the fit
within their errors.
The uncertainty related to the choice of a specific functional form for the
shape of the $\mathcal{NN}$ distribution is evaluated by using
a binned PDF as an alternative description. All relative and absolute normalizations
of $B$ background components that are fixed in the fit are varied within their errors.
For all these variations, the corresponding change in the fitted signal yield is
taken as a systematic uncertainty.

The branching fractions are calculated from the measured signal yields
assuming
$\BR(\Upsilon (4S)\to\BzBzb) = \BR(\Upsilon (4S)\to\BpBm) = 0.5$.
The results are listed in Table~\ref{tab:results}.
For $\bomg$, we also compute the $90\%$ confidence level
(C.L.) upper limit
$\BR(\Bz\to\omega\gamma) < \BFomUL\times10^{-6}$
using a Bayesian technique.
We determine the
branching fraction upper limit ${\cal B}_l$
such that
$\int_0^{{\cal B}_l}{\calL}\,d{\cal B}/\int_0^\infty {\calL}\, d{\cal B} = 0.90$, assuming
a flat prior in the branching fraction
and taking into account the systematic uncertainty.

We  test the hypothesis of isospin symmetry by measuring
the quantity $\Gamma(\brpg)/[2\Gamma(\brzg)] - 1 = \IST $.
The result is consistent  with the theoretical expectation~\cite{ali2004}.

The isospin-averaged branching fraction is extracted from a simultaneous
fit to the three decay modes:
\begin{equation}\label{eq:res1}
  \avbr= (\BFav) \times10^{-6}.
\end{equation}
In the fit we impose the isospin constraints on the widths of the decay modes:
$ \Gamma_{B\to\rho^+\gamma}=2\Gamma_{B\to\rho^0\gamma}=2\Gamma_{B\to\omega \gamma}$.
Our measurements of the individual branching fractions are consistent with this 
hypothesis with a $\chi^2$ of 1.8 for 2 degrees of freedom. 
The significance of the signal is \significance $\sigma$, including
systematic uncertainties. This result is consistent with 
the measurement from  Belle~\cite{newbelle}. 
If we exclude the $\Bz\to\omega\gamma$ mode from the simultaneous fit, we obtain
$\BR(B\to\rho\gamma) = (\BFavrhorho )\times10^{-6}$.
Using the world average  value of \BR(\bkg) \cite{pdg}, we calculate
$\avbr/\BR(\bkg) = \BrhoBKst$.
%
This result can be used to calculate the ratio
$\VtdVts$ \cite{ali2004, Bosch:2004nd, Ball:2006nr}.
Following \cite{Ball:2006nr},
we obtain
\begin{equation}\label{eq:res5}
\VtdVts = \VtdVtsval ,
\end{equation}
where the first error is experimental and the second is theoretical.
This result is consistent with the
 measurement of this ratio from the study of $B^0$ and $B^0_s$ mixing~\cite{bsmixing}.

In conclusion, we have measured the branching fractions of
$\BR(\Bp\to\rhop\gamma) = (\BFrp)\times10^{-6}$
and
$\BR(\Bz\to\rhoz\gamma) = (\BFrz)\times10^{-6}$,  
and set a  $90\%$ C.L. upper limit on the $\bomg$ branching
fraction of
$\BR(\Bz\to\omega\gamma) < \BFomUL\times10^{-6}$.
The isospin-averaged branching fraction $\avbr= (\BFav) \times10^{-6}$ 
is  the most precise measurement of this quantity to date. 
This measurement is used to extract 
$\VtdVts = \VtdVtsval$.\\

We are grateful for the excellent luminosity and machine conditions
provided by our \pep2\ colleagues, 
and for the substantial dedicated effort from
the computing organizations that support \babar.
The collaborating institutions wish to thank 
SLAC for its support and kind hospitality. 
This work is supported by
DOE
and NSF (USA),
NSERC (Canada),
IHEP (China),
CEA and
CNRS-IN2P3
(France),
BMBF and DFG
(Germany),
INFN (Italy),
FOM (The Netherlands),
NFR (Norway),
MIST (Russia),
MEC (Spain), and
PPARC (United Kingdom). 
Individuals have received support from the
Marie Curie EIF (European Union) and
the A.~P.~Sloan Foundation.


\begin{thebibliography}{99}
\bibitem{ChargeConj}
Charge conjugate modes are 
implied throughout.
\end{thebibliography}
\end{document}